\documentclass{appolb}
\usepackage{graphicx}
\usepackage{cite}
\usepackage{url}
\usepackage{lineno}
\usepackage{wrapfig}
\usepackage{setspace,lipsum}
\usepackage[belowskip=-5pt,aboveskip=0pt]{caption}
\begin{document}
\title{Measurements of Fourier harmonics of azimuthal anisotropy in Pb+Pb and Xe+Xe collisions.%
\thanks{Presented at XIII Workshop on Particle Correlations and Femtoscopy}%
}
\author{Klaudia Burka
\address{Institute of Nuclear Physics Polish Academy of Sciences\\ul. Radzikowskiego 152, 31-342 Krakow, Poland}
\\
{}
}
\maketitle

\vspace{-1cm}
\begin{abstract}
The high-statistics data sets collected by the LHC experiments during the heavy-ion runs allow for a detailed study of the azimuthal anisotropy of charged particles.
This report presents the 
selected results of $v_{n}$ measured in Pb+Pb collisions at $\sqrt{s_{\mathrm{NN}}} = 5.02$ TeV and Xe+Xe collisions at $\sqrt{s_\mathrm{NN}} = 5.44$ TeV, with emphasis on the higher order Fourier coefficients, $v_{n}$, sensitive to initial state fluctuations of the medium created in ultra-relativistic heavy-ion collisions.
\end{abstract}
\vspace{-0.3cm}
\PACS{24.85}

 \vspace{-0.5cm}
\section{Introduction}

Strongly-interacting Quark-Gluon Plasma (QGP), produced in ultra-relativistic heavy-ion collisions, exhibits hydrodynamical properties such as collective flow~\cite{Ollitrault:1992bk,Voloshin:2008dg}.\
The initial geometry of the non-central heavy-ion collision determine the initial elliptical shape of the interaction region.\
The dynamics within the plasma, as it expands, translate the spatial asymmetry of the initial-state into the final-state anisotropy in the momentum space.\
 This leads to an anisotropic particle distribution, commonly expressed with the Fourier series: $\frac{dN}{d\phi} \propto 1 + \sum_{n}2v_{n}\mathrm{cos}\left[n\left(\phi - \Psi_{n}\right)\right]$, where $\phi$ is the particle's azimuthal angle, $\Psi_{n}$ is the azimuthal angle of the reaction plane of the overlap region and the $v_{n}$ is the Fourier coefficient of $n$-th order.\
The second order flow coefficient, $v_{2}$, called elliptic flow, is related to the spatial asymmetry of the collision zone.\
Higher order Fourier components of the angular distribution arise from fluctuations in the initial positions of the nucleons within nucleus, which is different from event to event.\

This report focuses on the recent measurements of $v_{n}$ harmonics obtained by the  ATLAS experiment in $\sqrt{s_{\mathrm{NN}}}\!=\!5.02$ TeV Pb+Pb~\cite{Aaboud:2018ves} and $\sqrt{s_{\mathrm{NN}}}\!=\!5.44$~TeV Xe+Xe collisions~\cite{ATLAS-CONF-2018-011}.\
Corresponding results from the \mbox{ALICE} and CMS experiments can be found in Refs.~\cite{Acharya:2018ihu,Acharya:2018lmh,CMS-PAS-HIN-18-001,Sirunyan:2017pan}.\
The Xe ion is almost twice smaller than the Pb ion, which introduces larger initial spatial fluctuations of the collision zone in the smaller system.\ 
Therefore, studying the Xe+Xe collisions fills the gap between the azimuthal anisotropy measurements done for small pp~\cite{Aad:2015gqa} and p+Pb~\cite{Aad:2013fja} and large Pb+Pb~\cite{ATLAS:2012at} systems.
 
Simultaneous measurements of the elliptic flow and higher order flow harmonics provide better understanding of the properties of the initial state interaction region and the QGP medium created in heavy-ion collisions.\
Over the years, several experimental techniques were developed to calculate the $v_{n}$ coefficients.\
Comparisons of $v_{n}$ harmonics measured with different methods directly probe flow harmonics fluctuations.\
Three commonly used methods are briefly described in the next section.

 \vspace{-0.35cm}
\section{The $v_{n}$ measurement}

The two-particle correlation (2PC) method~\cite{PhysRevC.44.1091} is based on measuring pair distributions in the relative pseudorapidity, $\Delta\eta\!=\eta_{a}\!-\eta_{b}$, and the azimuthal angle $\Delta\phi\!=\!\phi_{a}\! -\phi_{b}$ of correlated particles.\ 
In this technique, the $v_{n}$ harmonics can be extracted by parametrizing the two-particle azimuthal distribution
\begin{equation}
\frac{dN^{pair}}{d\Delta\phi} \propto 1 + \sum_{i=1}^{\infty}2v_{n}^{a}v_{n}^{b}cos(n\Delta\phi).
\end{equation}
The correlated particles are selected from given $p_{\mathrm{T}}$ ranges for each, the ``reference'' particle (``a'') and the ''associated'' particle (``b'').\ 
The 2PC method is focused on studying long-range correlations, thus the $\Delta\eta$ is usually chosen to be $|\Delta\eta| > 2$, to suppress non-flow correlations.\ Non-flow correlations are correlations due to processes not related to the flow phenomena, such as particle decays or jet production.

In the scalar-product (SP) method~\cite{Adler:2002pu} the $v_{n}$ coefficients are obtained using the scalar-products of a particle unit flow vector, $u\!=e^{in\phi}$ with the reference flow vectors, $Q_{n}\!=\sum_{j}w_{j}e^{in\phi_{j}}$, measured at forward rapidities.\
The 2PC and SP methods measure the same quantity, which is $\sqrt{\langle v_{n}^{2}\rangle}$~\cite{Luzum:2012da}.\ 
Any non-flow contributions in the SP method are suppressed by requiring a large separation between pseudorapidities of particles used for the unit and reference flow vectors determination, e.g.\ in ATLAS it is required that $|\Delta\eta| > 3.2$~\cite{Aaboud:2018ves}.\ 

The 2PC method can be extended to the
multi-particle cumulants of 2$k$-particle azimuthal correlations~\cite{Borghini:2000sa}.\ 
Commonly, the  $v_{n}$ harmonics are measured using 4-, 6- and 8- particle correlations.\
The cumulants of large number of particles (e.g.\ 8-particle cumulant) do not contain correlations of fewer number of particles (e.g.\ 2-particle correlations), thus, any non-flow effects are significantly suppressed.\
Additional advantage of the cumulant method is its sensitivity to flow fluctuations~\cite{Aad:2014vba}.\
Assuming Gaussian fluctuations, the 4-, 6- and 8-particle cumulant harmonics can be expressed by $v_{n}\{2k\}\! =\! \sqrt{v_{n}^{2} - \sigma_{v_{n}}^{2}}$ (for $k>$1), while the 2-particle cumulant harmonics $v_{n}\{2\}\! =\! \sqrt{v_{n}^{2} + \sigma_{v_{n}}^{2}}$, where the $ \sigma_{v_{n}}^{2}$ is the standard deviation of the $v_{n}$ distribution.\

\section{Results}
\subsection{Pb+Pb $v_{n}$ results}

\begin{figure}[h!]
\centering
\includegraphics[width=0.6\textwidth]{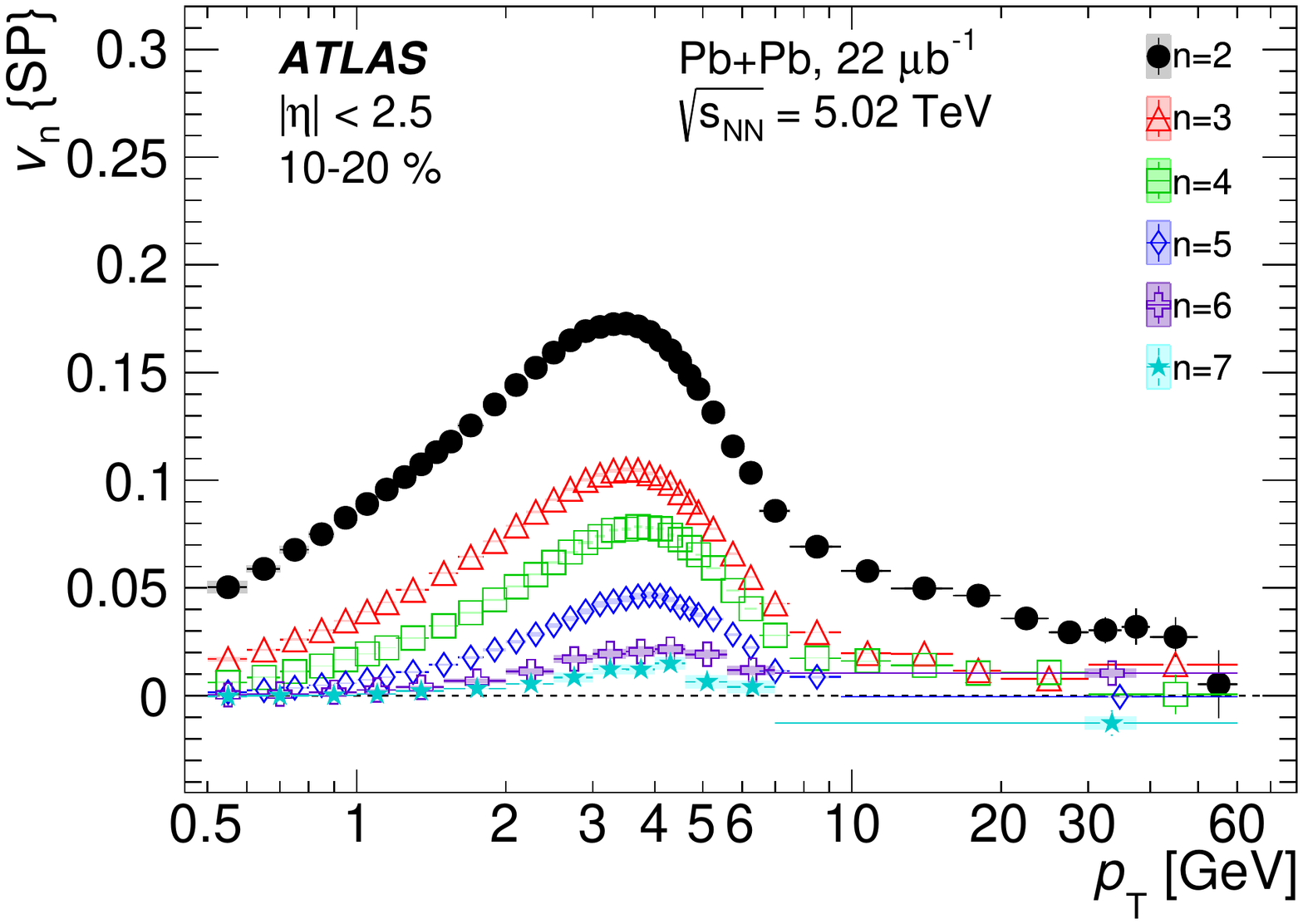}
  \caption{The $v_{n}(p_{\mathrm{T}})$ integrated over $|\eta|$~$<$ 2.5 in 10--20\% centrality interval obtained with the SP method in Pb+Pb collisions at $\sqrt{s_{\mathrm{NN}}}$ = 5.02~TeV~\cite{Aaboud:2018ves}.}
 \label{fig:PbPbvnSPpt}
\end{figure}

Fig.~\ref{fig:PbPbvnSPpt} shows the measurement of the $v_{n}\{\mathrm{SP}\}$ as a function of $p_{\mathrm{T}}$  in Pb+Pb collisions at $\sqrt{s_{\mathrm{NN}}}$~=~5.02~TeV in 10--20\% centrality interval~\cite{Aaboud:2018ves}.\
The observed $p_{\mathrm{T}}$ dependence is typical for $v_{n}$ harmonics.\
 An almost linear increase of $v_{n}$ values up to $p_{\mathrm{T}}\!=2$--3~GeV is measured
which is followed by a gradual rise to a maximum at $p_{\mathrm{T}}\!=3$--4~GeV and then continuous decrease for higher $p_{\mathrm{T}}$ is observed.\ 
The $v_{2}(p_{\mathrm{T}})$ values persist to be positive even at the highest measured $p_{\mathrm{T}}\!=60$~GeV,
which implies the path-length dependence
of parton energy loss in the created medium.\
 
\begin{figure}[h!]
\centering
\includegraphics[width=0.6\textwidth]{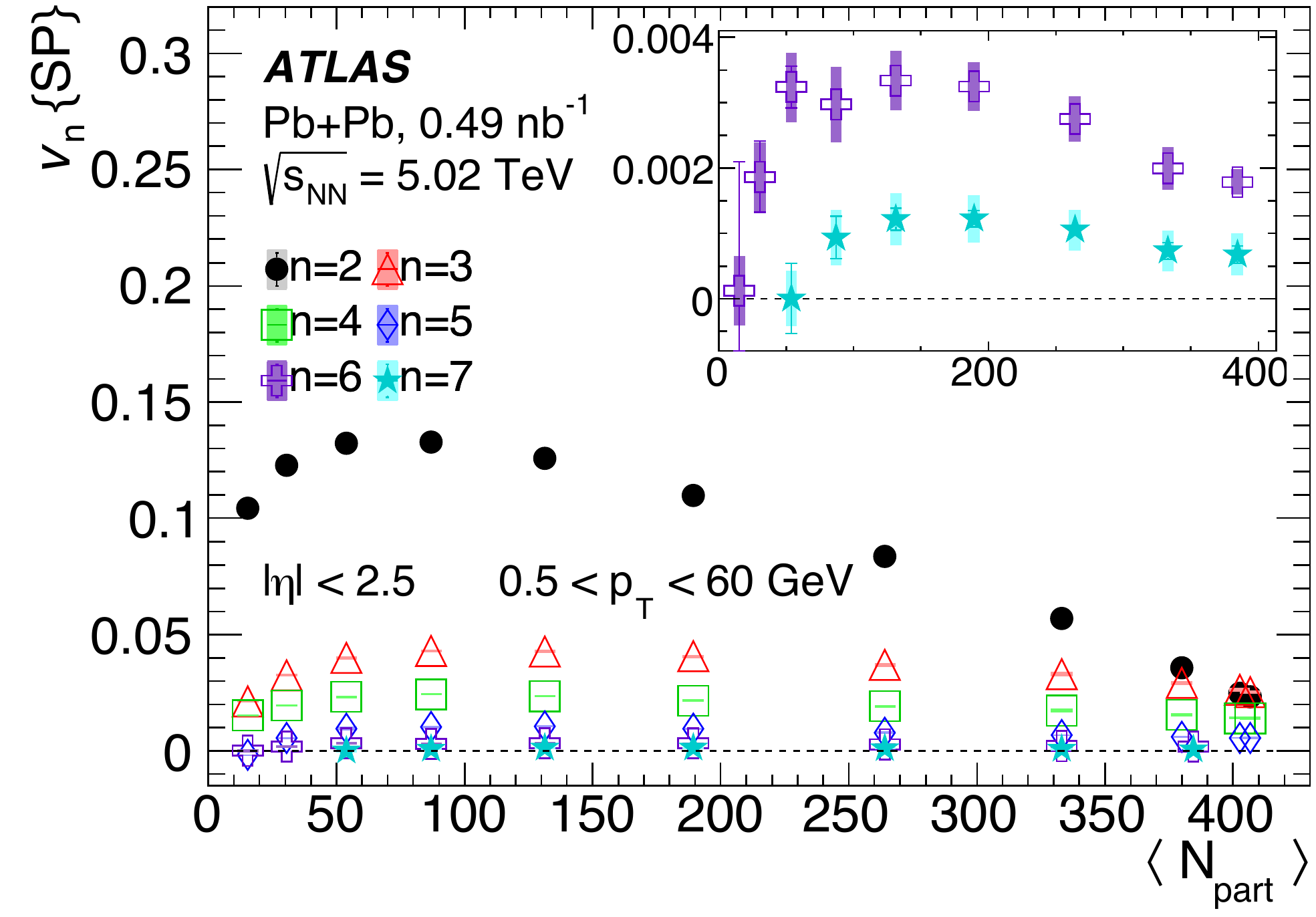}
\caption{Integrated $v_{n}$ vs.\ $N_{\mathrm{part}}$, integrated over $|\eta|<2.5$ and $p_{\mathrm{T}}$~= 0.5--60~GeV measured using the SP method in Pb+Pb collisions at $\sqrt{s_{\mathrm{NN}}} = 5.02$~TeV~\cite{Aaboud:2018ves}.\ The inset panel shows $v_{6}$ and $v_{7}$ with adjusted scale.}
\label{fig:PbPbvnSPcent}
\end{figure}

The dependence on the number of nucleons participating in heavy-ion collision, $N_{part}$, of the integrated $v_{n}$ measured using SP method in Pb+Pb collisions at $\sqrt{s_{\mathrm{NN}}} = 5.02$~TeV is presented in Fig.~\ref{fig:PbPbvnSPcent}.\
The second order flow coefficient is dominant almost over the whole centrality range except for the most central collisions, where the $v_{3}$ and $v_{4}$ become more prominent than $v_{2}$.\
This indicates that in the most central collisions the magnitude of $v_{n}$ harmonics is driven by fluctuations in the initial geometry of the collision zone.\
The $v_{7}$ harmonic is found to be non-zero for collisions of 0--50\% centrality.\ 
The elliptic flow and higher order flow coefficients, \mbox{$v_{2}$--$v_{5}$}, are also obtained in ultra-central collisions (0--0.1\%) corresponding to $N_{\mathrm{part}}$~=~406, see Fig.~\ref{fig:PbPbvnSPcent}.

\subsection{Xe+Xe $v_{n}$ results}

\begin{figure}[h!]
\centering
\includegraphics[width=\textwidth]{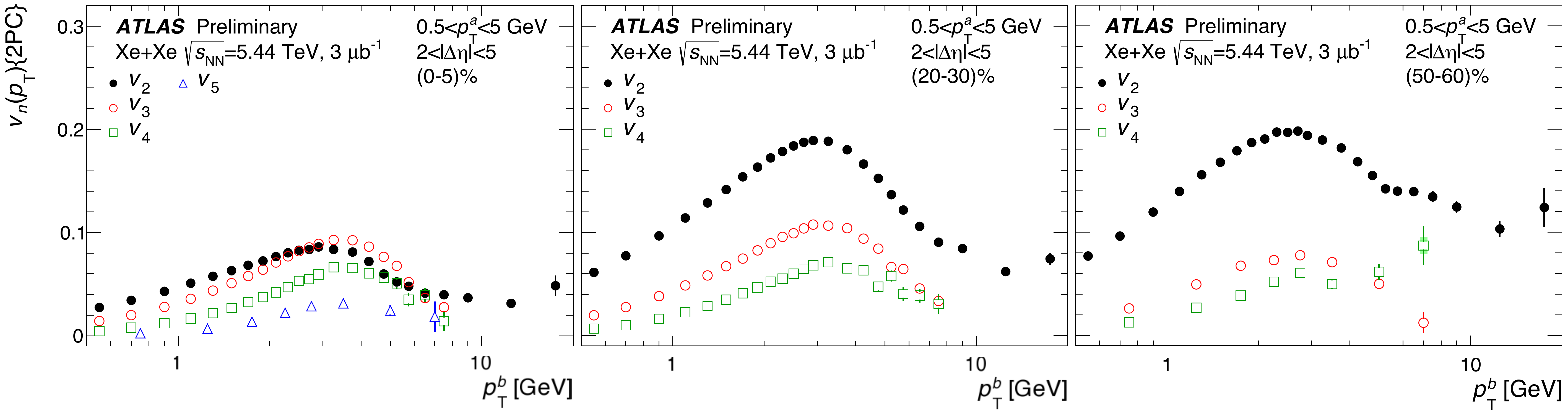}
\vspace{0.2cm}
\caption{The $v_{n}(p_{\mathrm{T}}^{\mathrm{b}})$ measured with the 2PC method in Xe+Xe collisions at $\sqrt{s_{\mathrm{NN}}} = 5.44$~TeV for $p_{\mathrm{T}}^{\mathrm{a}}$ = 0.5--5~GeV in three centrality intervals: 0--5\%, 20--30\% and 50--60\%~\cite{ATLAS-CONF-2018-011}.}
\label{fig:XeXevn2PCpt}
\end{figure}

The differential $v_{n}$ results as a function of $p_{\mathrm{T}}$ measured using 2PC method in Xe+Xe collisions at $\sqrt{s_{\mathrm{NN}}} = 5.44$~TeV are shown in Fig.~\ref{fig:XeXevn2PCpt}.\ The $v_{2}$--$v_{4}$ harmonics are measured for a wide range of $p_{\mathrm{T}}$ up to 20~GeV and are presented for three centrality intervals: 0--5\% (central collisions), 20--30\% (semi-central collisions) and 50--60\% (peripheral collisions).\ 
The $v_{5}$ coefficient is shown only for the most central collisions and up to $p_{\mathrm{T}} = 8$~GeV.\
The $v_{n}$ ordering observed in Xe+Xe collisions follow the same trend as observed in Pb+Pb collisions: $v_{2}\gg v_{3} > v_{4} > v_{5}$, and for the most central (0-5\%) collisions, where the $v_{3}$ becomes larger than $v_{2}$ for $p_{\mathrm{T}}> 3$~GeV.\

\begin{figure}[h!]
\centering
\includegraphics[width=0.32\textwidth]{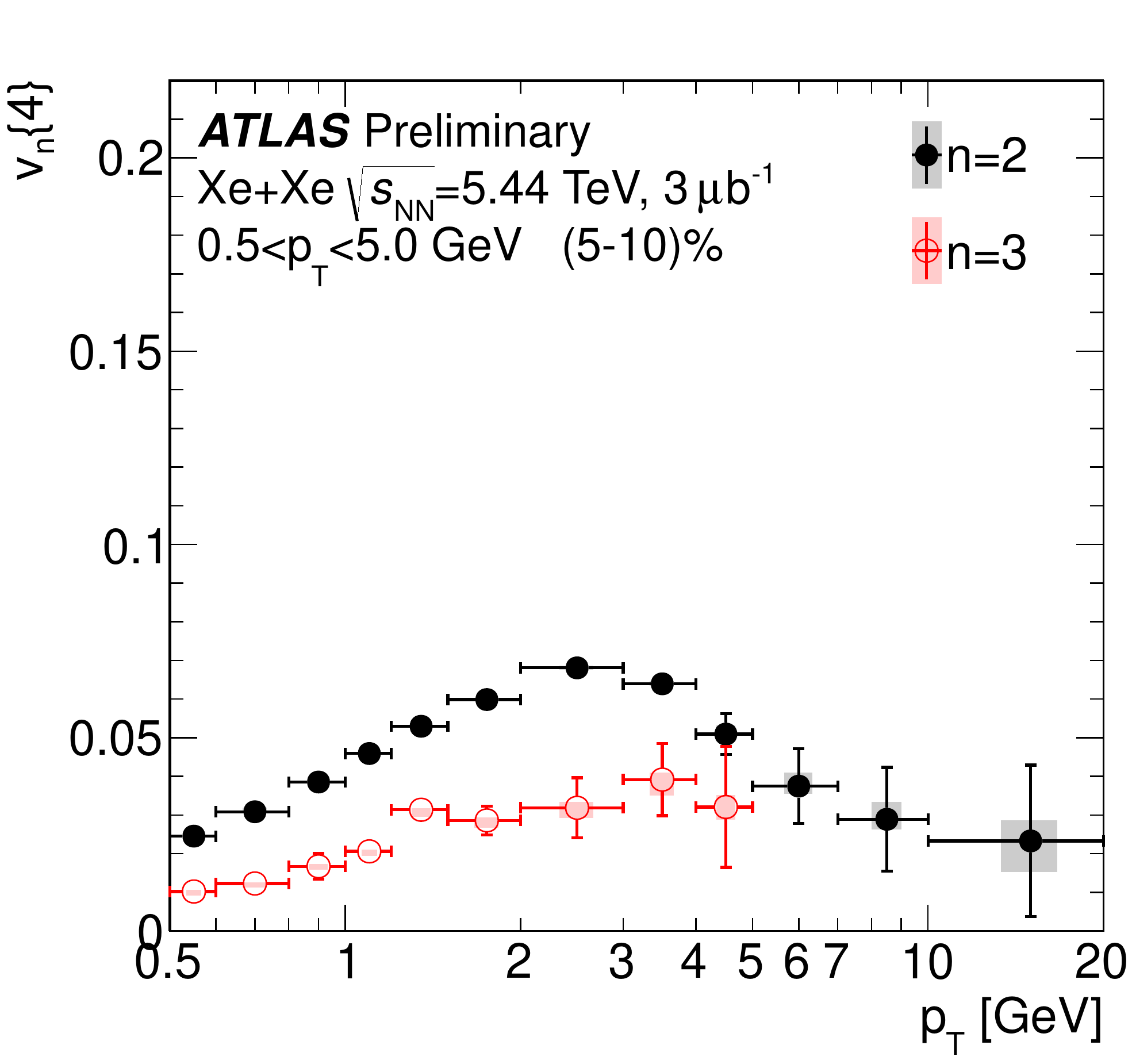}
\includegraphics[width=0.32\textwidth]{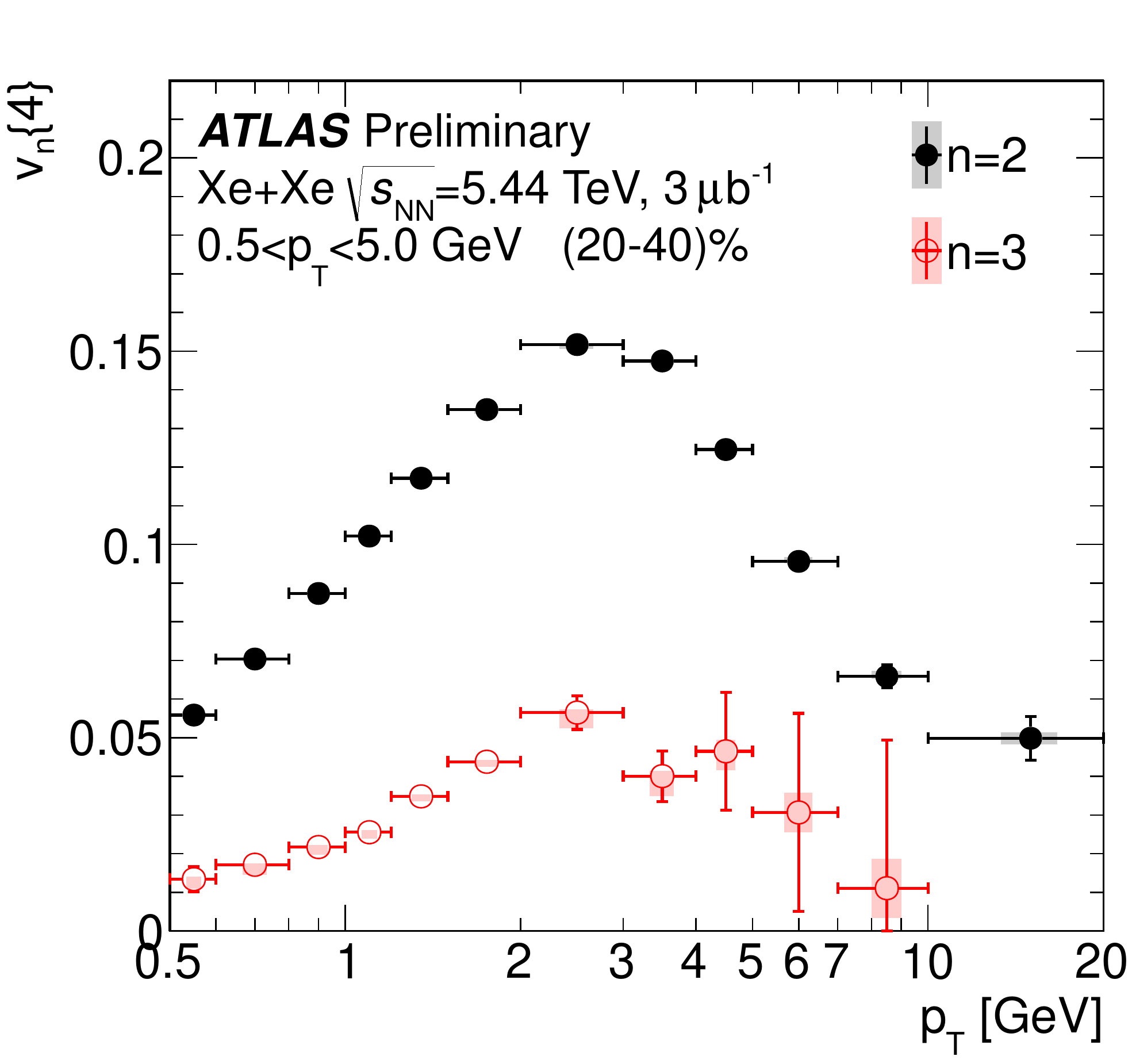}
\includegraphics[width=0.32\textwidth]{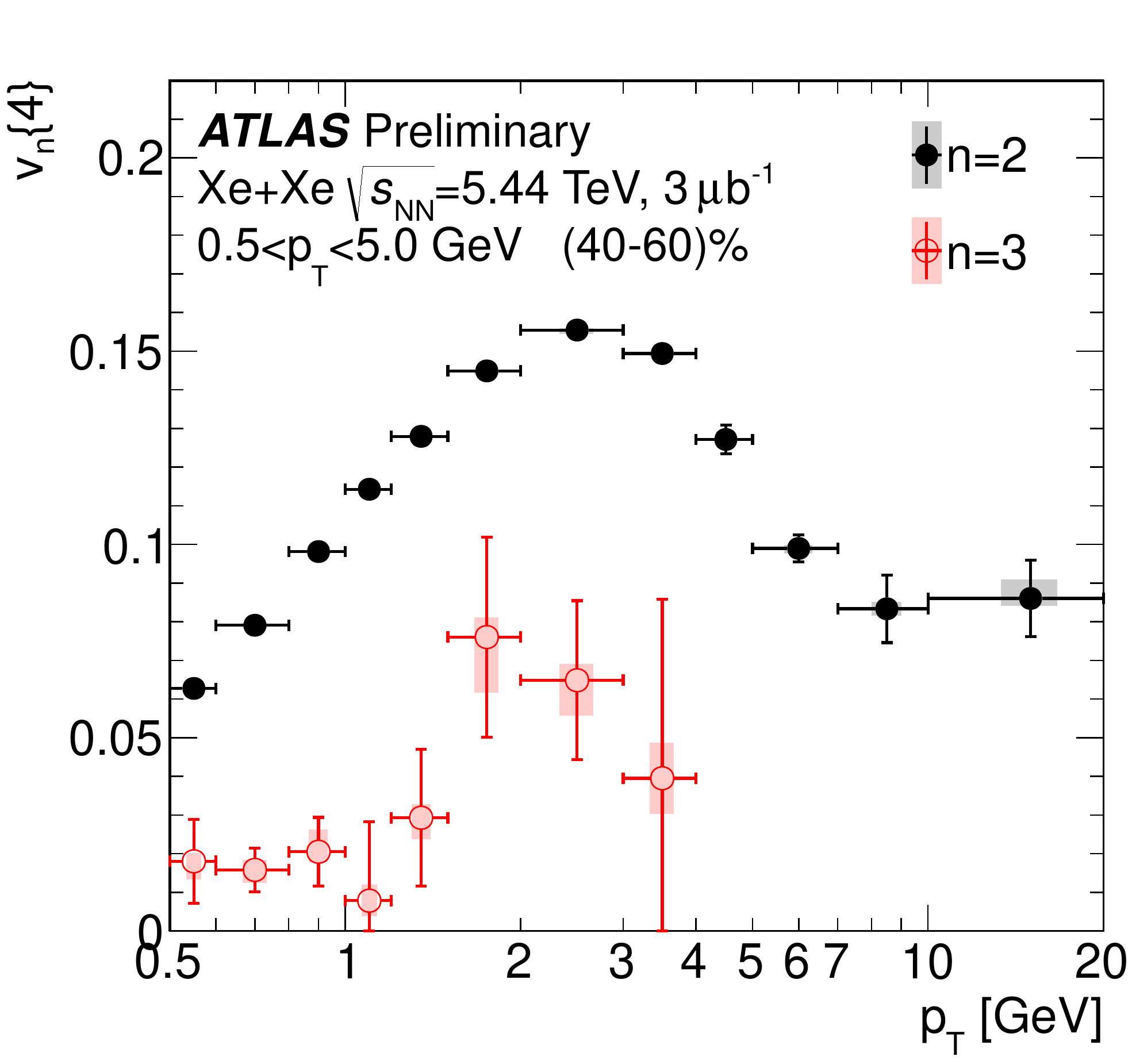}
\vspace{0.2cm}
\caption{\hspace{-0.1cm}The 4-particle cumulant $v_{n}\{4\}$ as a function of $p_{\mathrm{T}}$ measured in Xe+Xe collisions at $\sqrt{s_{\mathrm{NN}}} = 5.44$ TeV for three centrality intervals:\ 5--10\%, 20--40\% and 40--60\%~\cite{ATLAS-CONF-2018-011}}
\label{fig:XeXePbPb4c}
\end{figure}

The $v_{2}(p_{\mathrm{T}})$ and $v_{3}(p_{\mathrm{T}})$ obtained in Xe+Xe collisions with the 4-particle cumulant method are presented in Fig.~\ref{fig:XeXePbPb4c} for three centrality intervals: 5--10\%, 20--40\% and 40--60\%~\cite{ATLAS-CONF-2018-011}.\
The 4-particle cumulant is sensitive to $v_{n}$ fluctuations and suppresses two-particle non-flow correlations.\ 
Therefore, the $v_{n}\{4\}$ values are smaller than those obtained with the 2PC or SP methods, however similar trends in $p_{\mathrm{T}}$ and centrality dependences for cumulant, SP and 2PC results are observed.

\begin{figure}[h!]
\centering
\includegraphics[width=0.85\textwidth]{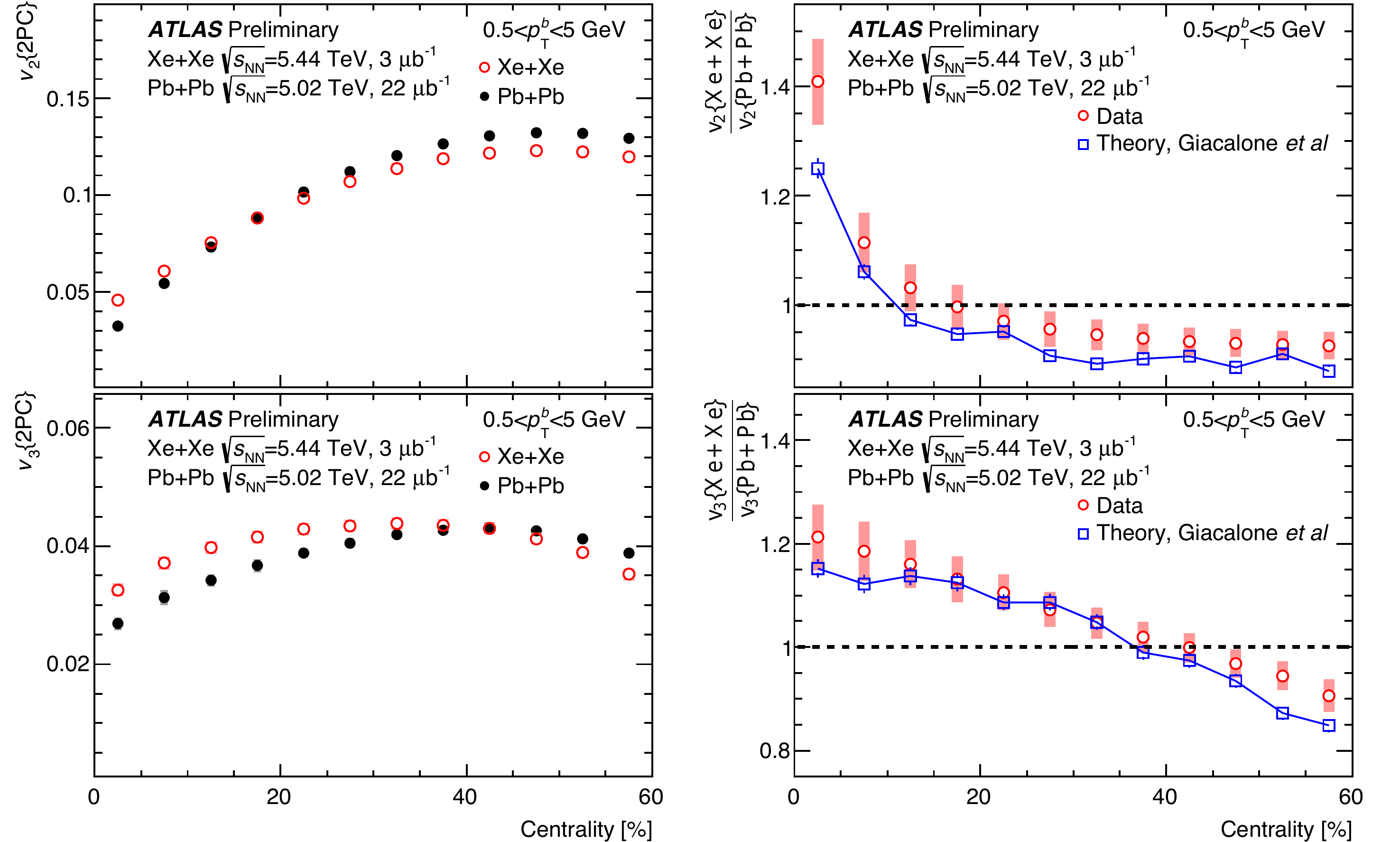}
\vspace{0.2cm}
\caption{\hspace{-0.1cm}Left panels:\ the centrality dependence of the $v_{2}$ (top) and $v_{3}$ (bottom) harmonics measured using 2PC method in Pb+Pb collisions at $\sqrt{s_{\mathrm{NN}}} = 5.02$~TeV and Xe+Xe collisions at $\sqrt{s_{\mathrm{NN}}} = 5.44$ TeV integrated over $0.5 < p_{\mathrm{T}} < 5 $ GeV~\cite{ATLAS-CONF-2018-011}.\
Right panels:\ The ratios of the Xe+Xe $v_{n}\{\mathrm{2PC}\}$ to the Pb+Pb $v_{n}\{\mathrm{2PC}\}$.\ The ratios are compared with theoretical predictions obtained for $p_{\mathrm{T}}$~= 0.3--5 GeV~\cite{Giacalone:2017dud}.
}
\label{fig:XeXePbPb2PC}
\end{figure}

The quantitative comparisons of the $v_{2}$ and $v_{3}$ between the Pb+Pb and Xe+Xe systems are shown in the left panels of Fig.~\ref{fig:XeXePbPb2PC}.\
 The $v_{n}$ harmonics, integrated over $p_{\mathrm{T}}$~= 0.5--5~GeV, obtained with the 2PC method are presented as a function of centrality percentiles.\  
  The ratios of  $v_{n}$\{Xe+Xe\} to $v_{n}$\{Pb+Pb\}, shown in the right panels,  are quite consistent with the theoretical predictions~\cite{Giacalone:2017dud}.\
The $v_{2}$\{Xe+Xe\} and $v_{3}$\{Xe+Xe\} values are significantly larger than for Pb+Pb collisions in the most central events.\ 
This is expected as 
in the smaller collision system the initial fluctuations are larger, which can modify the initial collision geometry,
 and thus, enhance the $v_{n}$.\
In the mid-central and peripheral collisions the $v_{n}$\{Xe+Xe\} is found to be systematically smaller than those in Pb+Pb collisions.\ 
This is expected from hydrodynamical models as the viscous effects are larger for the lighter, Xe+Xe, system.\
For the higher order flow coefficients, $v_{4}$ and $v_{5}$, the effect in most central events is less pronounced~\cite{ATLAS-CONF-2018-011}.

\section{Summary}
The LHC experiments provided many interesting results on the azimuthal anisotropy obtained with the 5.02 TeV Pb+Pb and 5.44 TeV Xe+Xe data using
the 2PC, SP and cumulant methods, which are sensitive to different aspects of heavy-ion collisions.
The $v_{n}$ harmonics show a similar $p_{\mathrm{T}}$ dependence: first an increase with $p_{\mathrm{T}}$ up to a maximum around 3--4~GeV and then a decrease for higher transverse momenta.
As a function of centrality, the $v_2$ harmonic significantly varies and dominates (but the most central collisions), while the higher order harmonics show a weak centrality dependence.
In the most central events, the Xe+Xe $v_2$ and $v_3$ values are larger than those in Pb+Pb collisions due to larger initial state fluctuations expected in the smaller system.
An opposite effect, observed in peripheral collisions, may indicate that the viscosity in the Xe+Xe is larger than the viscosity in Pb+Pb collisions. Significant fluctuations of $v_{n}$ harmonics, especially in central Xe+Xe collisions are also indicated by the four-particle cumulant results.
\\

\noindent This work was supported in part by the National Science Centre,\ Poland grants no.\ 2016/23/B/ST2/00702 and  2016/23/N/ST2/01339 and by PL-Grid Infrastructure.
\vspace{-0.7cm}
\begin{spacing}{0.9}
\bibliography{references}

\providecommand{\href}[2]{#2}\begingroup\raggedright\begin{thebibliography}{10}

\bibitem{Ollitrault:1992bk}
J.-Y. Ollitrault,
\href{http://dx.doi.org/10.1103/PhysRevD.46.229}{Phys. Rev. {\bfseries D46}
  (1992) 229--245}.

\bibitem{Voloshin:2008dg}
S.~A. Voloshin, A.~M. Poskanzer,  and R.~Snellings,
  \href{http://dx.doi.org/10.1007/978-3-642-01539-7_10}{Landolt-Bornstein
  {\bfseries 23} (2010) 293--333},
\href{http://arxiv.org/abs/0809.2949}{{\ttfamily arXiv:0809.2949 [nucl-ex]}}.

\bibitem{Aaboud:2018ves}
{ATLAS} Collaboration,
\href{http://arxiv.org/abs/1808.03951}{{\ttfamily arXiv:1808.03951 [nucl-ex]}}.

\bibitem{ATLAS-CONF-2018-011}
{ATLAS} Collaboration, Tech. Rep. ATLAS-CONF-2018-011.
\newblock \url{https://cds.cern.ch/record/2318870}.

\bibitem{Acharya:2018ihu}
{ALICE} Collaboration,
\href{http://dx.doi.org/10.1016/j.physletb.2018.06.059}{Phys. Lett. B
  {\bfseries 784} (2018) 82--95}.

\bibitem{Acharya:2018lmh}
{ALICE} Collaboration,
\href{http://dx.doi.org/10.1007/JHEP07(2018)103}{JHEP {\bfseries 07} (2018)
  103}.

\bibitem{CMS-PAS-HIN-18-001}
{CMS} Collaboration, Tech. Rep. CMS-PAS-HIN-18-001.
\newblock \url{https://cds.cern.ch/record/2318319}.

\bibitem{Sirunyan:2017pan}
{CMS} Collaboration,
\href{http://dx.doi.org/10.1016/j.physletb.2017.11.041}{Phys. Lett. B
  {\bfseries 776} (2018) 195--216}.

\bibitem{Aad:2015gqa}
{ATLAS} Collaboration,
\href{http://dx.doi.org/10.1103/PhysRevLett.116.172301}{Phys. Rev. Lett.
  {\bfseries 116} (2016) 172301}.

\bibitem{Aad:2013fja}
{ATLAS} Collaboration,
\href{http://dx.doi.org/10.1016/j.physletb.2013.06.057}{Phys. Lett. B
  {\bfseries 725} (2013) 60--78}.

\bibitem{ATLAS:2012at}
{ATLAS} Collaboration,
\href{http://dx.doi.org/10.1103/PhysRevC.86.014907}{Phys. Rev. C {\bfseries 86}
  (2012) 014907}.

\bibitem{PhysRevC.44.1091}
S.~Wang, Y.~Z. Jiang, Y.~M. Liu, D.~Keane, D.~Beavis, S.~Y. Chu, S.~Y. Fung,
  M.~Vient, C.~Hartnack,  and H.~St\"ocker,
  \href{http://dx.doi.org/10.1103/PhysRevC.44.1091}{Phys. Rev. C {\bfseries 44}
  (1991) 1091--1095}, \url{https://link.aps.org/doi/10.1103/PhysRevC.44.1091}.

\bibitem{Adler:2002pu}
{STAR} Collaboration,
\href{http://dx.doi.org/10.1103/PhysRevC.66.034904}{Phys. Rev. C {\bfseries 66}
  (2002) 034904}.

\bibitem{Luzum:2012da}
M.~Luzum and J.-Y. Ollitrault,
\href{http://dx.doi.org/10.1103/PhysRevC.87.044907}{Phys. Rev. C {\bfseries 87}
  (2013) 044907}.

\bibitem{Borghini:2000sa}
N.~Borghini, P.~M. Dinh,  and J.-Y. Ollitrault,
\href{http://dx.doi.org/10.1103/PhysRevC.63.054906}{Phys.\ Rev.\ C {\bfseries
  63} (2001) 054906}.

\bibitem{Aad:2014vba}
{ATLAS} Collaboration,
\href{http://dx.doi.org/10.1140/epjc/s10052-014-3157-z}{Eur. Phys. J. C
  {\bfseries 74} (2014) 3157}.

\bibitem{Giacalone:2017dud}
G.~Giacalone, J.~Noronha-Hostler, M.~Luzum,  and J.-Y. Ollitrault,
\href{http://dx.doi.org/10.1103/PhysRevC.97.034904}{Phys. Rev. C {\bfseries 97}
  (2018) 034904}.

\end{thebibliography}\endgroup
\bibliographystyle{BibStyleWoTitle}
\end{spacing}

\end{document}